\documentclass[twocolumn]{aastex63}
\usepackage{graphicx}
\usepackage{makecell}

\shorttitle{NuSTAR Spectroscopy of EXO 1846--031}
\shortauthors{Draghis et al.}

\begin{document}

\title{A New Spin on an Old Black Hole: NuSTAR Spectroscopy of EXO 1846--031}

\author[0000-0002-2218-2306]{Paul A. Draghis}
\email{pdraghis@umich.edu}
\affiliation{Department of Astronomy, University of Michigan, 1085 South University Avenue, Ann Arbor, MI 48109, USA}

\author{Jon M. Miller}
\affiliation{Department of Astronomy, University of Michigan, 1085 South University Avenue, Ann Arbor, MI 48109, USA}

\author{Edward M. Cackett}
\affiliation{Wayne State University, Department of Physics \& Astronomy, 666 W Hancock St, Detroit, MI 48201, USA}

\author{Elias S. Kammoun}
\affiliation{Department of Astronomy, University of Michigan, 1085 South University Avenue, Ann Arbor, MI 48109, USA}

\author{Mark T. Reynolds}
\affiliation{Department of Astronomy, University of Michigan, 1085 South University Avenue, Ann Arbor, MI 48109, USA}

\author{John A. Tomsick}
\affiliation{Space Sciences Laboratory, 7 Gauss Way, University of California, Berkeley, CA 94720-7450, USA}

\author{Abderahmen Zoghbi}
\affiliation{Department of Astronomy, University of Michigan, 1085 South University Avenue, Ann Arbor, MI 48109, USA}

\begin{abstract}
The black hole candidate EXO 1846-031 underwent an outburst in 2019, after at least 25 yr in quiescence. We observed the system using \textit{NuSTAR} on 2019 August 3.  The 3--79~keV spectrum shows strong relativistic reflection features. Our baseline model gives a nearly maximal black hole spin value of $a=0.997_{-0.002}^{+0.001}$ ($1\sigma$ statistical errors).  This high value nominally excludes the possibility of the central engine harboring a neutron star. Using several models, we test the robustness of our measurement to assumptions about the density of the accretion disk, the nature of the corona, the choice of disk continuum model, and the addition of reflection from the outer regions of the accretion disk. All tested models agree on a very high black hole spin value and a high value for the inclination of the inner accretion disk of $\theta\approx73^\circ$. We discuss the implications of this spin measurement in the population of stellar mass black holes with known spins, including LIGO and Virgo events.
\end{abstract}


\section{Introduction} \label{sec:intro}

Relativistic disk reflection (e.g., \citealt{2006ApJ...652.1028B}; \citealt{2015PhR...548....1M}) is a pragmatic tool for measuring black hole spin, in part because it requires no knowledge of the black hole mass and distance.  For these reasons, it is applicable across the black hole mass scale.  The main feature of the relativistic reflection spectrum is Fe K fluorescence, merely due to the high abundance and fluorescence yield of iron (\citealt{10.1093/mnras/249.2.352}).  If the gas is neutral, the line is predicted at 6.4~keV.  If the gas is ionized, the line is detected at progressively higher energies, up to 6.97 keV for H-like \ion{Fe}{26}.
The relativistic Doppler shifts and gravitational red-shifts experienced by gas orbiting in the innermost accretion disk cause the profile of the Fe K emission lines to be ``blurred'' (see \citealt{2000PASP..112.1145F}; \citealt{2003PhR...377..389R}; \citealt{Miller_2007} ).  Since the size of the Innermost Stable Circular Orbit (ISCO) decreases with increasing black hole spin (e.g., \citealt{1972ApJ...178..347B}; \citealt{novikov1973black}), the extremity of the blurring encodes the radius of the ISCO and therefore the spin of the black hole. 

A fundamental assumption in all measurements of black hole spin is that a standard, optically thick, geometrically thin accretion disk (\citealt{1973A&A....24..337S}) extends to the ISCO, and that any gas at smaller radii is on plunging orbits and is optically thin (\citealt{Reynolds_2008}).  In this case, there would be a clear dividing line between the region that is transparent to incident X-rays, and the optically thick disk that is reflective.  It is not currently possible to verify if real fluid disks obey the ISCO defined for test particles.  However, independent numerical simulations suggest that the necessary circumstance holds below a fraction of the Eddington limit (e.g., \citealt{Reynolds_2008}; \citealt{2008ApJ...676..549S}).  In recent observations of the black hole candidate MAXI J1820$+$070, \citealt{2020MNRAS.493.5389F} noted a small degree of extra-thermal emission that could originate at the inner edge of the plunging region, potentially consistent with non-zero torques.  However, even in this source, a stable and typical disk reflection spectrum is observed over considerable changes in flux (\citealt{2019MNRAS.490.1350B}), suggesting that the inner edge of the reflective disk is still determined by gravity.

Particularly in view of the growing number of binary black hole (BBH) mergers detected with LIGO and Virgo (see, e.g., \citealt{2019PhRvX...9c1040A}), it is important to determine the spin distribution of stellar-mass black holes with ordinary stellar companions.  Already, the distribution appears to be skewed toward very high spin values, perhaps indicative of formation in gamma-ray bursts (e.g., \citealt{2011ApJ...731L...5M}) or caused by an observational bias toward detection of high spin black holes (e.g., \citealt{2016MNRAS.458.2012V}).  This distribution may also point to solid-body rotation in black hole progenitor stars in order to explain the observed angular momenta, providing a rare but indirect insight into the most massive stars.  Comparing the black hole spin distribution in standard X-ray binaries to the distribution inferred from black holes in LIGO and Virgo events, both premerger and postmerger, can help to reveal relationships between these populations and the astrophysics underlying any significant disparities.  
Since the absolute number of stellar-mass black holes is still fairly small, every source contributes to these goals.  Especially for X-ray binaries that lie close to the Galactic plane, where line-of-sight obscuration may foil attempts to measure the black hole mass via optical techniques and spins via disk continuum measurements (e.g., \citealt{2006ApJ...652..518M}; \citealt{2008ApJ...679L..37L}; \citealt{2010ApJ...718L.122G}), disk reflection is the most viable way forward.

EXO 1846-031 was discovered on 1985 April 3 during a slew maneuver of the {\em EXOSAT} mission (\citealt{IAUC4051}).   Searches for the optical counterpart failed (\citealt{1985IAUC.4059....3W}).  Based on our fits, at least (see below), this is likely the result of a very high column along the line of sight. 
Later, based on its X-ray spectrum from the 1985 {\em EXOSAT} observation, composed of an ultra-soft component and a high-energy power law tail, \citealt{1993A&A...279..179P} argued that this object is a Low Mass X-ray Binary (LMXB) system that likely harbors a black hole.   A second outburst of EXO 1846$-$031 may have been observed with {\it CGRO/BATSE} (\citealt{IAUC6096}).  The source was not detected with any monitor since at least 1994.  However, on 2019 July 23rd, {\em MAXI} detected a hard X-ray transient consistent with the location of EXO 1846-031 (\citealt{2019ATel12968....1N}), indicating a new outburst.  The source was soon localized using the Neil Gehrels {\em Swift} observatory (\citealt{2019ATel12969....1M}).  This facilitated follow-up in radio bands: EXO 1846$-$031 was detected at 5.25 and 7.45~GHz (2.54 $\pm$ 0.03~mJy, and 2.42 $\pm$ 0.03~mJy, respectively) using the VLA (\citealt{2019ATel12977....1M}), and at 1.28~GHz with MeerKAT (\citealt{2019ATel12992....1W}).  

By virtue of its passband, sensitivity, and ability to observe very bright sources, {\em NuSTAR} (\citealt{2013ApJ...770..103H}) is the best observatory for studying disk reflection in black holes and neutron stars, and for constraining black hole spin and stellar radii in these systems (e.g., \citealt{King_2014}; \citealt{2016ApJ...826L..12E}; \citealt{2019ApJ...887..184T};  \citealt{2020ApJ...893...30X}).  Motivated by the opportunity to measure the fundamental properties of a stellar-mass black hole, we requested a TOO/DDT observation with {\em NuSTAR}.  In Section 2, we detail the data reduction and preparation that were undertaken.  Section 3 describes the spectral modeling and results.  In Section 4, we discuss the results in light of other recent efforts to measure black hole spin parameters.

\section{Observations and Data Reduction} \label{sec:obs}
\begin{figure}[ht]
\includegraphics[width=0.45\textwidth]{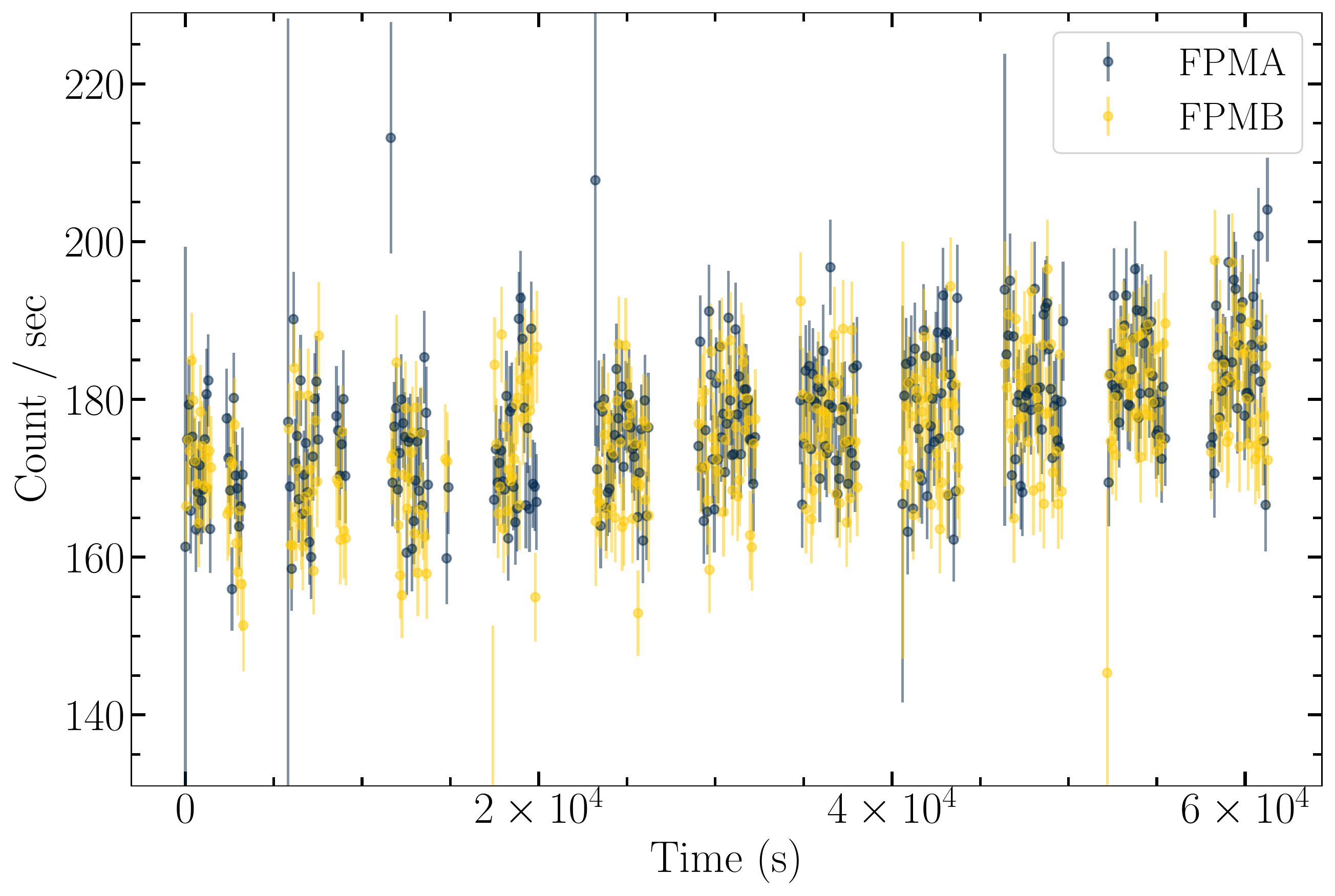}
\caption{Light curve of the observation. For clarity, only one-tenth of the points were plotted.}
\label{fig:light_curve}
\end{figure}

\textit{NuSTAR} observed EXO 1846-031 on 2019 August 3 starting at 02:01:09 UT, under ObsID 90501334002.  A net exposure of 22.2~ks was obtained, with an average of approximately 280 counts s$^{-1}$ when combining the FPMA and FPMB sensors.  The light curve in Figure \ref{fig:light_curve} shows a lack of strong variability in the count rate (e.g., no evidence of flares, dips, or state transitions), making it safe to characterize the source properties through fits to the time-averaged spectra.

The data were reduced using the routines in HEASOFT v6.26.1 through the NuSTARDAS pipeline v1.8.0 and CALDB v20190812.  Source spectra were extracted from circular regions (centered on the known source position) with radii of 180$^{\prime\prime}$ in the two FPM sensors.  Regions of the same size were used for extraction of background events.  The resulting spectra were grouped using the ``ftool'' \texttt{grppha}, such that each energy bin contains at least 30 counts.

\newpage
\section{Analysis and Results} \label{sec:analysis}
\begin{figure}[ht]
\includegraphics[width=0.49\textwidth]{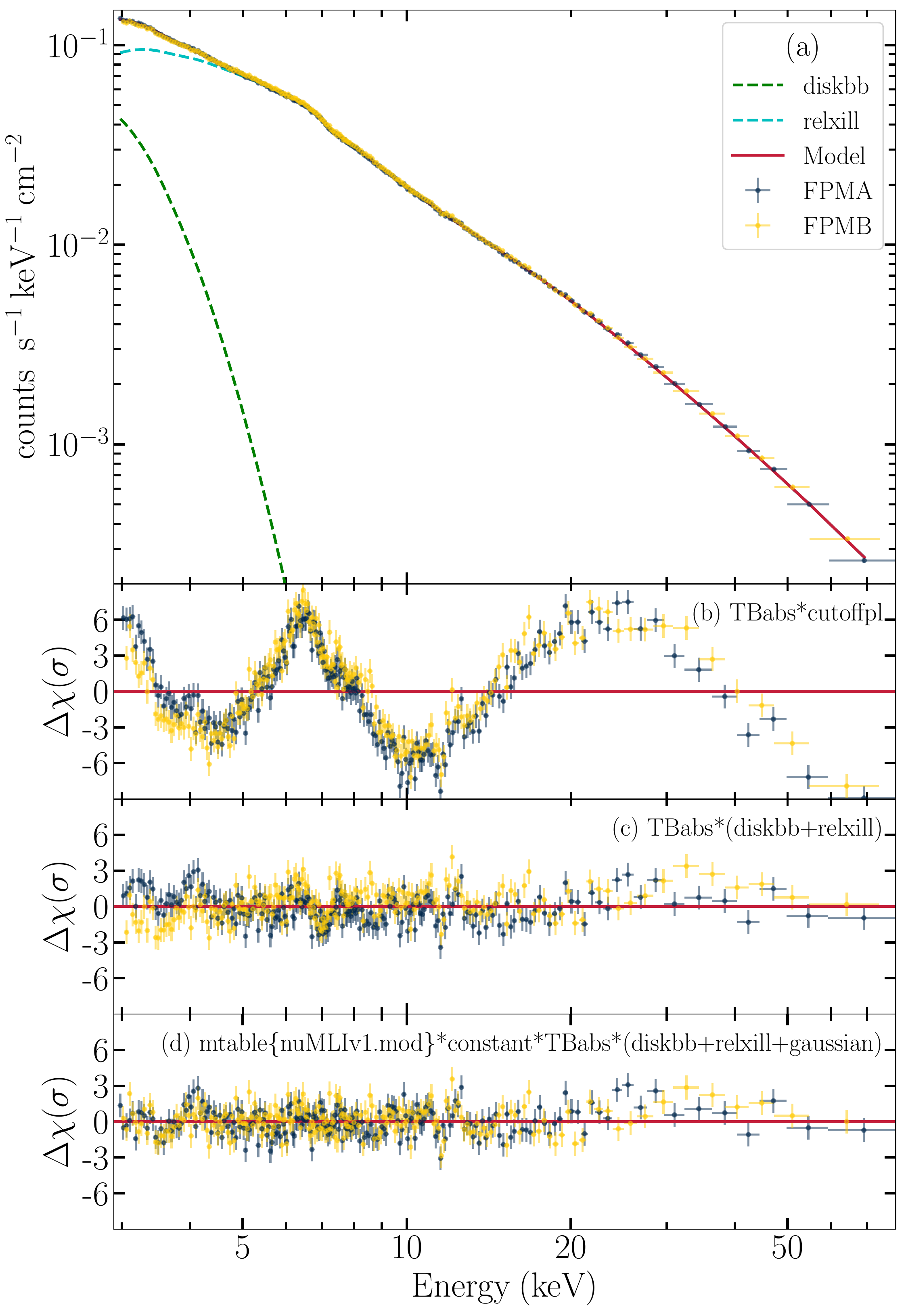}
\caption{Panel (a) shows the spectrum of EXO 1846-031. The blue points represent the data from the FPMA sensor of NuSTAR, and maize points are the data from the FPMB sensor. 
For model fitting, the spectra were grouped to require at least 30 counts per bin; additional binning was used in this plot for visual clarity only.
The red line shows the \texttt{mtable\{nuMLIv1.mod\}*constant*TBabs*(diskbb+relxill+\\gaussian)} model, with the green line representing the contribution of the \texttt{diskbb} component and the cyan line that the one of the \texttt{relxill} component.  Panel (b) shows the residuals in terms of sigma for the \texttt{TBabs*cutoffpl} model.  Panel (c) shows the residuals for the \texttt{TBabs*(diskbb+relxill)} model, and panel (d) is for \texttt{mtable\{nuMLIv1.mod\}*constant*TBabs*(diskbb+relxill+\\gaussian)}, which we regard as our ``baseline'' model.}
\label{fig:fit}
\end{figure}

All spectral fitting was performed using XSPEC v12.10.1n (\citealt{1996ASPC..101...17A}).  The fits were made across the full \textit{NuSTAR} energy band (3--79 keV).  The FPMA and FPMB spectra (top panel of Figure \ref{fig:fit}) were fit jointly.    The quality of the fit was measured using $\chi^{2}$ statistics, appropriate to the high signal in each bin. 

As a first step, we fit the spectra with an absorbed cutoff power law (\texttt{TBabs*cutoffpl}); a very poor fit resulted ($\chi^2/\nu$=8525.80/2601 = 3.28, where $\nu$ is the numbers of degrees of freedom).  This initial attempt returns a power law index of $\Gamma$=2.04 and a high-energy cutoff at $\sim$200 keV, but strong residuals remain  (see panel (b) in Figure \ref{fig:fit}).  The unmodeled features are consistent with a strong relativistic disk reflection, including a Fe K line complex and a broad flux excess in the 20--40 keV range.  The strength and breadth of the residuals suggest that the disk likely extends to the ISCO, or at least to very small radii.

In order to account for the reflection features, the spectra were next fit with \texttt{relxill} v1.2.1 (\citealt{2013MNRAS.430.1694D}; \citealt{2014ApJ...782...76G}).   This model is the standard in recent efforts to measure relativistic disk reflection, and is powerful in that it now has variants that enable examinations of the corona, the density of the gas in the accretion disk, and more.  The cutoff power law model also failed to capture some low-energy flux; this could be the effect of the $\chi^{2}$ minimization balancing positive and negative residuals, but it could represent continuum emission from the disk reaching above 3~keV.  Therefore, we also included a \texttt{diskbb} component (\citealt{1984PASJ...36..741M}), to describe any weak emission from the disk above 3~keV.  The interstellar absorption was accounted for through the multiplicative component \texttt{TBabs} (\citealt{2006HEAD....9.1360W}), using the abundances described by \citealt{2000ApJ...542..914W}.  

The \texttt{relxill} component parameters include the spin of the black hole, $a = cJ/GM^{2}$ ($-1 \leq a \leq 1$), the inclination at which the reflector is viewed (here, it is important to note that this is the inclination of the {\em inner} disk, not necessarily the binary system), the power law index of the illuminating flux ($\Gamma$), the cutoff energy of the power law ($E_{\rm cut}$), the ionization of the reflector ($\xi$ \footnote{$\xi = L/nr^{2}$, where $L$ is the luminosity of the source, $n$ is the hydrogen number density of the reflector, and $r$ is the distance between the source and reflector.}), and the iron abundance of the reflector ($A_{\rm Fe}$, measured relative to the solar value).  The \texttt{relxill} model can also nominally measure the inner and outer radius of the accretion disk, independently of the spin parameter.  In all fits, we found that the inner radius was consistent with the ISCO, so we fixed $r_{\rm in} = 1.0~r_{\rm ISCO}$ in this and all other fits with \texttt{relxill}.  Similarly, we fixed the outer radius to a value just shy of its maximum ($r_{\rm out} = 990~\rm GM/c^{2}$, whereas the nominal maximum value is $1000~\rm GM/c^{2}$) in order to avoid numerical artifacts.  In this basic version of \texttt{relxill}, the geometry of the corona is not prescribed, and the emissivity is treated as a broken power law: $J \propto r^{-q_1}$ for $r < r_{\rm break}$, and $J\propto r^{-q_2}$ for $r > r_{\rm break}$.  This is a simplified mathematical form that may eventually be supervened by very sensitive spectra, but it broadly adheres to ray-tracing studies of the emissivity close to spinning black holes (e.g., \citealt{2020MNRAS.493.5532W}).  Finally, \texttt{relxill} also includes a ``reflection fraction'' parameter.  In other reflection models (e.g., \texttt{pexmon}, \citealt{2007MNRAS.382..194N}), this parameter has a geometric interpretation, but in \texttt{relxill} this is a scale factor that, in this case, proved to be degenerate with the component normalization, so it was fixed to unity.  The \texttt{diskbb} component has only two parameters, the color temperature of the disk ($kT$) and a flux normalization.  It is important to note that the \texttt{diskbb} component does not include a zero-torque inner boundary condition (\citealt{2005ApJ...618..832Z}).  The \texttt{TBabs} component has only one parameter, the equivalent neutral hydrogen column density ($N_{\rm H}$).  

For this object, the line-of-sight column density is so high ($\geq 10^{23}$) that most of the spectrum below 2-3~keV is unavailable. Due to this, any observations using the \textit{Swift} X-ray telescope (XRT; \citealt{2007SPIE.6686E..07B}) or the \textit{Neutron star Interior Composition Explorer Mission (NICER)} X-ray telescope (\citealt{2012SPIE.8443E..13G}) would not offer any advantage in extending the \textit{NuSTAR} passband.

This model yielded a vastly improved fit over the cutoff power law: $\chi^2/\nu$=2897.29/2592=1.12.  The residuals are shown in panel (c) of Figure \ref{fig:fit}.   The reflection spectrum has clearly been modeled extremely well, but a negative flux residual is evident at approximately 7~keV in both the FPMA and FPMB sensors.  Disk winds are seen in the spectra of stellar-mass black holes viewed at moderate and high inclinations, when the disk is important.  For instance, a qualitatively similar feature was seen in 4U 1630$-$472 (\citealt{King_2014}).  In that case, {\it NuSTAR} was also able to effectively separate and measure the disk reflection and disk wind features.  To model the feature in EXO 1846$-$031, a simple additive Gaussian absorption line was introduced.  This produced an improvement of $\Delta\chi^2=49$ through the addition of three extra parameters, indicating that the Gaussian absorption line is significant at the $6\sigma$ level of confidence.

Additionally, the spectrum shows a difference between the two FPM \textit{NuSTAR} sensors at low energies. This increase in flux in FPMA is caused by a tear in its thermal blanket, described by \citealt{2020arXiv200500569M}. To account for this difference, we used the prescription described by \citealt{2020arXiv200500569M}. We fit the spectra allowing a constant offset between FPMA and FPMB spectra, ignoring the 3--7~keV band of FPMA. Upon fixing the multiplicative constant between the two sensors and including the 3--7~keV FPMA signal, we added a multiplicative table component provided by \citealt{2020arXiv200500569M}, fixing the MLI covering fraction of FPMB to 1 and allowing the covering fraction of FPMA to vary. Fitting the complete model again for the entire 3--79~keV \textit{NuSTAR} band gives a covering fraction of FPMA of $\sim 0.88$. {\em We regard this as our baseline model}, and in XSPEC parlance it can be written as \texttt{mtable\{nuMLIv1.mod\}*constant*TBabs*\\(diskbb+relxill+gaussian)}.  The residuals of the fit using this model are shown in panel (d) of Figure \ref{fig:fit} and the model in relation to the spectra in panel (a).  This fit produces $\chi^2/\nu$=2726.24/2588=1.05.  Even upon this addition to the model, the Gaussian component remains significant at the $6\sigma$ level of confidence.  This improvement to the model does not influence any of the parameter estimates.

The ``best fit'' parameter values were determined through $\chi^{2}$ minimization.  These predictions were used for initial values for the prior distributions in MCMC chains \footnote{The XSPEC EMCEE implementation written by Zoghbi A. was used: https://zoghbi-a.github.io/xspec\_emcee/ }.  The chains were run with 100 walkers for $10^5$ steps, with an equally long burn-in, and trimmed to exclude the chains that obviously did not converge to a solution. The trimmed chains were used as a proposal distribution for a new set of chains, which were run for a total of $3\times10^6$ steps, with no burn-in.  This approach was used in order to avoid introducing any bias in the proposal distribution for the samples.  The values presented in this paper represent medians  across the chain samples, and the errors represent the $\pm 1\sigma$ limits on the distributions of each parameter in the chain. These errors are purely statistical, with the systematic errors likely being much larger and primarily driven by uncertainties in whether or not the disk adheres to the test particle ISCO.  While running the $\chi^2$ minimization, the parameters of the Gaussian absorption line around 7 keV were allowed to vary.  Due to the lack of degeneracy between the parameters of the Gaussian line and the other parameters of the models, for simplicity, for the MCMC chains they were fixed to the best fit value.  The $\chi^2$ and degrees of freedom reported in this paper are the values obtained from $\chi^2$ minimization, including the free parameters for the \texttt{gaussian} and  \texttt{mtable\{nuMLIv1.mod\}*constant} components.  The fit parameters of this model are shown in column 1 of Table \ref{models}, and the model compared to the data in panels (a) and (d) in Figure \ref{fig:fit}. For reference, the corner plot of the parameters of the \texttt{relxill} component in model 1 is given in Appendix \ref{sec:app}.  Importantly, there are no obvious degeneracies between the model parameters.  

The continuum parameters achieved with the baseline model suggest that we observed EXO 1846$-$031 in a hard intermediate state.  The disk temperature is fairly low, $kT = 0.43\pm 0.01~\rm keV$, projecting only a small amount of flux into the {\it NuSTAR} band.  The power law index is moderate, $\Gamma = 2.00\pm 0.01$.  The high cutoff energy, $E_{\rm cut} = 198^{+5}_{-6}$~keV, is typical of such states and suggests that a combination of thermal and non-thermal processes may power the corona.  Non-thermal processes such as magnetic heating enable the corona to be compact.  This is consistent with the high value of the inner emissivity that is also measured in the baseline model ($q_1 = 7\pm 1$), based on models showing that strong light bending and time delays result from compact coronae close to rapidly spinning black holes (e.g., \citealt{2012MNRAS.424.1284W}; however, see \citealt{2012A&A...545A.106S} and \citealt{2019MNRAS.485..239K}).  It is also interesting to note that the model does not require an enhanced iron abundance ($A_{\rm Fe} = 0.86\pm 0.05$).  Some fits to black hole spectra with this variant of \texttt{relxill} require highly elevated abundances; this is potentially the result of modeling with a disk density that is too low (e.g., \citealt{2018ApJ...855....3T}).  

\begin{figure}[ht]
\includegraphics[width=0.49\textwidth]{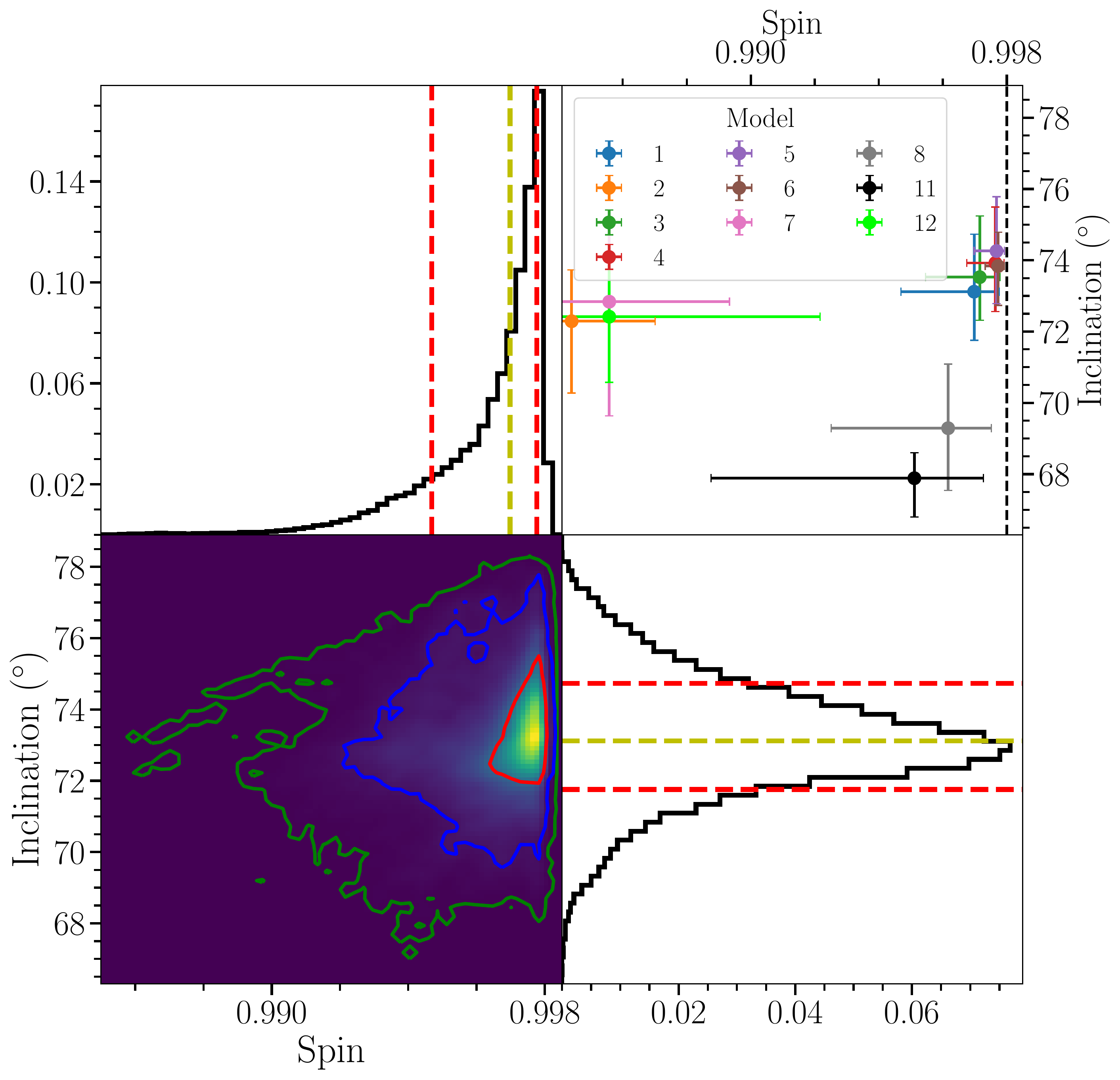}
\caption{Histogram of the spin-inclination parameter space for the \texttt{mtable\{nuMLIv1.mod\}*constant*TBabs*(diskbb+\\relxill+gaussian)} model. In the lower left panel, the red, blue, and green contours represent lines of $1\sigma$, $2\sigma$, and $3\sigma$ . The top left and lower right panels show the 1D histogram of the parameters. The yellow line represents the median of the distribution and the red lines represent the $\pm1\sigma$ confidence limits of the median.
The top right panel represents the spin-inclination parameter space for the models presented in Table \ref{models}. The black dashed line shows the maximum theoretical prediction for the spin of a black hole with a thin disk of 0.998 (\citealt{1974ApJ...191..507T}). Models 9 and 10 are omitted. }
\label{fig:hist}
\end{figure}

Most importantly, the baseline model predicts a near maximal and well-constrained spin parameter: $a=0.997 _{- 0.002 }^{+ 0.001 }$.  The inclination of the inner disk is also very well constrained, to $ 73^{\circ}\pm 2^{\circ}$.   These parameters can be partly degenerate, since both gravitational red-shifts and {\it relativistic} Doppler shifts create line asymmetry.  They are not fully degenerate since it is the extremity of gravitational red-shifting that drives the spin determination, and the location of the blue horn of the Fe K fluorescence complex that drives the inclination measurement. The bottom left panel of Figure \ref{fig:hist} shows the $1\sigma$, $2\sigma$, and $3\sigma$ confidence intervals (in red, blue, and green) on the spin measurement of our baseline model, with relation to the inclination of the inner disk. The 1D histograms show the posterior samples of the two parameters in the MCMC chain. The distribution of the samples for the spin falls almost entirely above 0.990 and the mode of the distribution indicates an even higher value than median of the $3\times10^6$ samples. There appears to be no degeneracy between the two parameters.

\begin{splitdeluxetable*}{cccccccccBccc|cc|cc}
\tablecaption{Results \label{models}}
\tablewidth{300pt} 
\tabletypesize{\scriptsize}
\tablehead{
\colhead{Model} & \colhead{$\#1$. relxill} & 
\colhead{$\#2$. relxill} & \colhead{$\#3$. relxill} & 
\colhead{$\#4$. relxillD} & \colhead{$\#5$. relxillD} & 
\colhead{$\#6$. relxillD} & \colhead{$\#7$. relxill}&
\colhead{$\#8$. relxillCp} &
\colhead{Model} & \colhead{$\#9$. relxilllp} & 
\colhead{$\#$10. relxilllpCp} &
\colhead{Model} & \colhead{$\#11$. relxill}& 
\colhead{Model} & \colhead{$\#12$. reflionx$\_$hd}\\ 
\colhead{Parameters} & \colhead{} & \colhead{($q_2=3$)} & \colhead{($R_{\rm br}=6$)} & 
\colhead{(n=15)} & \colhead{(n=17)} & \colhead{(n=19)} &
\colhead{+xillver} & \colhead{} & 
\colhead{Parameters} & 
\colhead{} &  \colhead{}&
\colhead{Parameters} & \colhead{+kerrbb}&
\colhead{Parameters} & \colhead{+diskbb}
}
\startdata\\
  $cf_A$ &
  $ 0.884 _{- 0.009 }^{+ 0.009 }*$ &
  $ 0.883 _{- 0.009 }^{+ 0.008 }*$ &
  $ 0.883 _{- 0.008 }^{+ 0.009 }*$ &
  $ 0.884 _{- 0.013 }^{+ 0.009 }*$ &
  $ 0.884 _{- 0.012 }^{+ 0.009 }*$ &
  $ 0.883 _{- 0.009 }^{+ 0.009 }*$ &
  $ 0.887 _{- 0.009 }^{+ 0.008 }*$ &
  $ 0.884 _{- 0.009 }^{+ 0.009 }*$ &
  $cf_A$ &
  $ 0.872 _{- 0.009 }^{+ 0.009 }*$ &
  $ 0.884 _{- 0.009 }^{+ 0.009 }*$ &
  $cf_A$ &
  $ 0.881 _{- 0.009 }^{+ 0.009 }*$ &
  $cf_A$ &
  $ 0.888 _{- 0.009 }^{+ 0.009 }*$\\
  $\rm const_A $ &
  $ 0.991 _{- 0.001 }^{+ 0.001 }*$ &
  $ 0.991 _{- 0.001 }^{+ 0.001 }*$ &
  $ 0.991 _{- 0.001 }^{+ 0.001 }*$ &
  $ 0.991 _{- 0.001 }^{+ 0.001 }*$ &
  $ 0.991 _{- 0.002 }^{+ 0.001 }*$ &
  $ 0.991 _{- 0.001 }^{+ 0.001 }*$ &
  $ 0.992 _{- 0.001 }^{+ 0.001 }*$ &
  $ 0.991 _{- 0.001 }^{+ 0.001 }*$ &
  $\rm const_A $ &
  $ 0.990 _{- 0.001 }^{+ 0.001 }*$ &
  $ 0.991 _{- 0.001 }^{+ 0.001 }*$ &
  $\rm const_A $ &
  $ 0.991 _{- 0.001 }^{+ 0.001 }*$&
  $\rm const_A $ &
  $ 0.992 _{- 0.001 }^{+ 0.001 }*$ \\
  $N_{\rm H} \: (\rm \times 10^{22} \: cm^{-2})$ &
  $ 11.0 _{- 0.5 }^{+ 0.5 }$ &
  $ 11.5 _{- 0.4 }^{+ 0.4 }$ &
  $ 11.0 _{- 0.5 }^{+ 0.5 }$ &
  $ 11.9 _{- 0.5 }^{+ 0.5 }$ &
  $ 11.8 _{- 0.5 }^{+ 0.5 }$ &
  $ 11.5 _{- 0.5 }^{+ 0.5 }$ &
  $ 9.3 _{- 0.5 }^{+ 0.5 }$ &
  $ 10.8 _{- 0.5 }^{+ 0.5 }$ &
  $N_{\rm H} \: (\rm \times 10^{22} \: cm^{-2})$ &
  $ 11.1 _{- 0.5 }^{+ 0.5 }$ &
  $ 8.1 _{- 0.5 }^{+ 0.8 }$ &
  $N_{\rm H} \: (\rm \times 10^{22} \: cm^{-2})$ &
  $ 11.5 _{- 0.4 }^{+ 0.3 }$ &
  $N_{\rm H} \: (\rm \times 10^{22} \: cm^{-2})$ &
  $ 12.6 _{- 0.4 }^{+ 0.5 }$ \\
  $kT_{\rm in} \: (\rm keV)$ &
  $ 0.43 _{- 0.01 }^{+ 0.01 }$ &
  $ 0.44 _{- 0.01 }^{+ 0.01 }$ &
  $ 0.43 _{- 0.01 }^{+ 0.01 }$ &
  $ 0.41 _{- 0.01 }^{+ 0.01 }$ &
  $ 0.41 _{- 0.01 }^{+ 0.01 }$ &
  $ 0.42 _{- 0.01 }^{+ 0.01 }$ &
  $ 0.42 _{- 0.01 }^{+ 0.01 }$ &
  $ 0.42 _{- 0.01 }^{+ 0.01 }$ &
  $kT_{\rm in} \: (\rm keV)$ &
  $ 0.46 _{- 0.01 }^{+ 0.01 }$ &
  $ 0.41 _{- 0.02 }^{+ 0.02 }$ &
  $\eta$ &
  $ 0.2 _{- 0.1 }^{+ 0.2 }$ &
  $kT_{\rm in} \: (\rm keV)$ &
  $ 0.40 _{- 0.01 }^{+ 0.01 }$\\
  $\rm norm_{\rm d} \: (\rm \times 10^{4})$ &
  $ 3.9 _{- 0.6 }^{+ 0.7 }$ &
  $ 3.5 _{- 0.4 }^{+ 0.4 }$ &
  $ 3.9 _{- 0.7 }^{+ 0.6 }$ &
  $ 6.2 _{- 0.8 }^{+ 0.7 }$ &
  $ 5.9 _{- 0.8 }^{+ 0.7 }$ &
  $ 4.9 _{- 0.6 }^{+ 0.6 }$ &
  $ 3.1 _{- 0.5 }^{+ 0.7 }$ &
  $ 4.2 _{- 0.6 }^{+ 0.7 }$ &
  $\rm norm_{\rm d} \: (\rm \times 10^{4})$ &
  $ 2.4 _{- 0.3 }^{+ 0.4 }$ &
  $ 2.7 _{- 0.5 }^{+ 1.0 }$ &
  \makecell{$M_{\rm bh} (\rm M_\odot)$\\  $M_{\rm dd} \: (\rm 10^{15}g/s)$} &
  \makecell{$ 9 _{- 4 }^{+ 5 }$\\  $ 2 _{- 1 }^{+ 3 }$} &
  $\rm norm_{\rm d} \: (\rm \times 10^{4})$ &
  $ 9.1 _{- 0.9 }^{+ 1.3 }$ \\
  $q_1$ &
  $ 7.3 _{- 0.8 }^{+ 0.8 }$ &
  $ 9.5 _{- 0.9 }^{+ 0.4 }$ &
  $ 7.5 _{- 0.7 }^{+ 0.8 }$ &
  $ 7.4 _{- 0.7 }^{+ 0.7 }$ &
  $ 7.8 _{- 0.8 }^{+ 0.7 }$ &
  $ 9.0 _{- 0.7 }^{+ 0.5 }$ &
  $ 9.0 _{- 1.5 }^{+ 0.8 }$ &
  $ 5.9 _{- 0.5 }^{+ 0.6 }$ &
  \nodata &
  \nodata &
  \nodata &
  \makecell{$D_{\rm bh} \: (\rm kpc) $\\  $f_{\rm col} $} &
  \makecell{$7*$ \\  $ 1.62 _{- 0.04 }^{+ 0.04 }$} &
  $\rm norm_{\rm cutoffpl}$ &
  $ 3.69 _{- 0.04 }^{+ 0.05 }$\\
  $q_2$ &
  $ 0.7 _{- 0.4 }^{+ 0.4 }$ &
  $3*$ &
  $ 1.0 _{- 0.3 }^{+ 0.2 }$ &
  $ 1.1 _{- 0.6 }^{+ 0.4 }$ &
  $ 0.7 _{- 0.5 }^{+ 0.5 }$ &
  $ 0.5 _{- 0.4 }^{+ 0.5 }$ &
  $q_1*$ &
  $ 0.6 _{- 0.4 }^{+ 0.4 }$ &
  \nodata &
  \nodata &
  \nodata &
  \makecell{$q_1$\\  $q_2$} &
  \makecell{$ 5.2 _{- 0.5 }^{+ 0.4 }$\\  $ 0.9 _{- 0.4 }^{+ 0.4 }$} &
  \makecell{$q_1$\\  $q_2$} &
  \makecell{$ 6.8 _{- 1.3 }^{+ 2.7 }$\\  $ 0.6 _{- 0.5 }^{+ 0.8 }$}  \\
  $R_{\rm br} \: (\rm r_g)$ &
  $ 8 _{- 2 }^{+ 3 }$ &
  $ 9 _{- 3 }^{+ 3 }$ &
  $6*$ &
  $ 6 _{- 2 }^{+ 4 }$ &
  $ 8 _{- 2 }^{+ 3 }$ &
  $ 7 _{- 2 }^{+ 2 }$ &
  $10*$ &
  $ 13_{- 4 }^{+ 5 }$&
  $h \: (\rm r_g)$ &
  $ 24 _{- 4 }^{+ 6 }$ &
  $ 69 _{- 19 }^{+ 18 }$ &
  $R_{\rm br} \: (\rm r_g)$ &
  $ 16 _{- 5 }^{+ 7 }$ &
  $R_{\rm br} \: (\rm r_g)$ &
  $ 12 _{- 3 }^{+ 8 }$\\
  $a$ &
  $ 0.997 _{- 0.002 }^{+ 0.001 }$ &
  $ 0.984 _{- 0.005 }^{+ 0.003 }$ &
  $ 0.997 _{- 0.002 }^{+ 0.001 }$ &
  $ 0.998 _{- 0.001 }$ &
  $ 0.998 _{- 0.001 }$ &
  $ 0.998 _{- 0.001 }$ &
  $ 0.986 _{- 0.008 }^{+ 0.004 }$ &
  $ 0.996 _{- 0.004 }^{+ 0.001 }$ &
  $a$ &
  $ 0.2 _{- 0.6 }^{+ 0.5 }$ &
  $ 0.1 _{- 0.7 }^{+ 0.7 }$ &
  $a$ &
  $ 0.995 _{- 0.006 }^{+ 0.002 }$ &
  $a$ &
  $ 0.986 _{- 0.003 }^{+ 0.007 }$\\
  $\rm Incl \: (^\circ)$ &
  $ 73 _{- 1 }^{+ 2 }$ &
  $ 72 _{- 2 }^{+ 1 }$ &
  $ 74 _{- 1 }^{+ 2 }$ &
  $ 74 _{- 1 }^{+ 2 }$ &
  $ 74 _{- 2 }^{+ 2 }$ &
  $ 74 _{- 1 }^{+ 1 }$ &
  $ 73 _{- 3 }^{+ 2 }$ &
  $ 69 _{- 2 }^{+ 2 }$ &
  $\rm Incl \: (^\circ)$ &
  $ 63 _{- 2 }^{+ 2 }$ &
  $ 72 _{- 1 }^{+ 1 }$ &
  $\rm Incl \: (^\circ)$ &
  $ 68 _{- 1 }^{+ 1 }$ &
  $\rm Incl \: (^\circ)$ &
  $ 72 _{- 2 }^{+ 2 }$\\
  $\Gamma$ &
  $ 2.00 _{- 0.01 }^{+ 0.01 }$ &
  $ 1.99 _{- 0.01 }^{+ 0.01 }$ &
  $ 2.00 _{- 0.01 }^{+ 0.01 }$ &
  $ 2.10 _{- 0.01 }^{+ 0.01 }$ &
  $ 2.10 _{- 0.01 }^{+ 0.01 }$ &
  $ 2.09 _{- 0.01 }^{+ 0.01 }$ &
  $ 2.03 _{- 0.02 }^{+ 0.02 }$ &
  $ 2.05 _{- 0.01 }^{+ 0.01 }$ &
  $\Gamma$ &
  $ 1.97 _{- 0.01 }^{+ 0.01 }$ &
  $ 2.01 _{- 0.01 }^{+ 0.01 }$ &
  $\Gamma$ &
  $ 1.98 _{- 0.01 }^{+ 0.01 }$ &
  $\Gamma$ &
  $ 2.30 _{- 0.01 }^{+ 0.01 }$ \\
  $\rm Log(\xi)$ &
  $ 3.48 _{- 0.03 }^{+ 0.04 }$ &
  $ 3.54 _{- 0.05 }^{+ 0.05 }$ &
  $ 3.47 _{- 0.03 }^{+ 0.04 }$ &
  $ 3.34 _{- 0.02 }^{+ 0.03 }$ &
  $ 3.31 _{- 0.04 }^{+ 0.03 }$ &
  $ 3.03 _{- 0.04 }^{+ 0.04 }$ &
  $ 3.52 _{- 0.07 }^{+ 0.07 }$ ($ 1.0 _{- 0.7 }^{+ 0.9 }$) &
  $ 3.53 _{- 0.04 }^{+ 0.05 }$ &
  $\rm Log(\xi)$ &
  $ 3.63 _{- 0.03 }^{+ 0.03 }$ &
  $ 3.54 _{- 0.03 }^{+ 0.04 }$ &
  $\rm Log(\xi)$ &
  $ 3.62 _{- 0.07 }^{+ 0.04 }$ &
  $\rm Log(\xi)$ &
  $ 2.21 _{- 0.07 }^{+ 0.09 }$\\
  $A_{\rm Fe}$ &
  $ 0.86 _{- 0.05 }^{+ 0.05 }$ &
  $ 0.87 _{- 0.05 }^{+ 0.05 }$ &
  $ 0.86 _{- 0.06 }^{+ 0.05 }$ &
  $ 0.62 _{- 0.05 }^{+ 0.05 }$ &
  $ 0.58 _{- 0.04 }^{+ 0.04 }$ &
  $ 0.62 _{- 0.02 }^{+ 0.02 }$ &
  $ 0.71 _{- 0.04 }^{+ 0.04 }$ &
  $ 0.74 _{- 0.04 }^{+ 0.04 }$ &
  $A_{\rm Fe}$ &
  $1*$ &
  $1*$ &
  $A_{\rm Fe}$ &
  $1*$ &
  $A_{\rm Fe}$ &
  $1*$ \\
  $E_{\rm cut} / kT_e \:(\rm keV)$ &
  $ 198 _{- 6 }^{+ 5 }$ &
  $ 200 _{- 5 }^{+ 5 }$ &
  $ 198 _{- 6 }^{+ 5 }$ &
  $300*$ &
  $300*$ &
  $300*$ &
  $ 190 _{- 12 }^{+ 10 }$ &
  $ 225 _{- 40 }^{+ 61 }$ &
  $E_{\rm cut}  / kT_e\:(\rm keV)$ &
  $ 132 _{- 5 }^{+ 6 }$ &
  $ 37 _{- 2 }^{+ 3 }$ &
  $E_{\rm cut}  \:(\rm keV)$ &
  $ 196 _{- 8 }^{+ 6 }$ &
  $E_{\rm cut} \:(\rm keV)$ &
  $300*$ \\
  $\rm Log(n)$ &
  $15*$ &
  $15*$ &
  $15*$ &
  $15*$ &
  $17*$ &
  $19*$ &
  $15*$ &
  $15*$ &
  $\rm Log(n)$ &
  $15*$ &
  $15*$ &
  $\rm Log(n)$ &
  $15*$ &
  $\rm Log(n)$ &
  $ 20.0 _{- 0.7 }^{+ 0.2 }$\\
  $\rm norm_{\rm r} \: (\rm \times 10^{-3})$ &
  $ 9.2 _{- 0.2 }^{+ 0.2 }$ &
  $ 9.0 _{- 0.1 }^{+ 0.2 }$ &
  $ 9.1 _{- 0.2 }^{+ 0.2 }$ &
  $ 11.9 _{- 0.5 }^{+ 0.5 }$ &
  $ 12.2 _{- 0.5 }^{+ 0.4 }$ &
  $ 12.4 _{- 0.6 }^{+ 0.5 }$ &
  $ 7.6 _{- 0.4 }^{+ 0.4 }$ ($ 7 _{- 1 }^{+ 2 }$) &
  $ 9.5 _{- 0.2 }^{+ 0.2 }$ &
  $\rm norm_{\rm r} \: (\rm \times 10^{-3})$ &
  $ 21.4 _{- 0.5 }^{+ 0.7 }$ &
  $ 21.08 _{- 0.3 }^{+ 0.4 }$ &
  $\rm norm_{\rm r} \: (\rm \times 10^{-3})$ &
  $ 9.2 _{- 0.2 }^{+ 0.2 }$ &
  $\rm norm_{\rm reflionx} \: (\rm \times 10^{-3})$ &
  $ 82 _{- 7 }^{+ 21 }$\\
  $E_{\rm g} \:(\rm keV)$ &
  $6.99_{-0.03}^{+0.03}*$ &
  $7.03_{-0.04}^{+0.08}*$ &
  $6.98_{-0.03}^{+0.03}*$ &
  $6.97_{-0.03}^{+0.03}*$ &
  $6.96_{-0.03}^{+0.03}*$ &
  $6.95_{-0.02}^{+0.03}*$ &
  $7.08_{-0.04}^{+0.04}*$ &
  $6.97_{-0.04}^{+0.04}*$ &
  $E_{\rm g} \:(\rm keV)$ &
  $7.04_{-0.03}^{+0.04}*$ &
  $7.09_{-0.04}^{+0.03}*$ &
  $E_{\rm g} \:(\rm keV)$ &
  $6.99_{-0.03}^{+0.03}*$ &
  $E_{\rm g} \:(\rm keV)$ &
  $6.92_{-0.04}^{+0.02}*$\\
  $\sigma_{\rm g} \: (\rm \times 10^{-2} \: keV)$ &
  $8_{-8}^{+4}*$ &
  $7_{-7}^{+8}*$ &
  $8_{-8}^{+5}*$ &
  $5_{-5}^{+6}*$ &
  $4_{-4}^{+6}*$ &
  $9_{-5}^{+4}*$ &
  $1_{-1}^{+10}*$ &
  $10_{-7}^{+4}*$ &
  $\sigma_{\rm g} \: (\rm \times 10^{-2} \: keV)$ &
  $4_{-3}^{+4}*$ &
  $16_{-3}^{+4}*$ &
  $\sigma_{\rm g} \: (\rm \times 10^{-2} \: keV)$ &
  $2_{-2}^{+6}*$ &
  $\sigma_{\rm g} \: (\rm \times 10^{-2} \: keV)$ &
  $13_{-3}^{+3}*$\\
  $norm_{\rm g} \: (\times 10^{-4})$ &
  $-5.0_{-0.6}^{+0.6}*$ &
  $-3.6_{-0.5}^{+0.5}*$ &
  $-5.0_{-0.7}^{+0.6}*$ &
  $-4.7_{-0.6}^{+0.6}*$ &
  $-4.5_{-0.6}^{+0.6}*$ &
  $-6.0_{-0.7}^{+0.6}*$ &
  $-2.7_{-0.5}^{+0.5}*$ &
  $-5.2_{-1.3}^{+0.9}*$ &
  $\rm norm_{\rm g} \: (\rm \times 10^{-4})$ &
  $-5.5_{-1.1}^{+1.1}*$ &
  $-7.3_{-0.7}^{+0.6}*$ &
  $\rm norm_{\rm g} \: (\rm \times 10^{-4})$ &
  $-4.3_{-0.6}^{+0.5}*$ &
  $\rm norm_{\rm g} \: (\rm \times 10^{-4})$ &
  $-9.4_{-0.3}^{+0.2}*$\\ 
  \tableline
  $\chi^2 / \nu$ &
  \makecell{$2726/2588$ \\ ($1.05$)} &
  \makecell{$2788/2589$ \\ ($1.08$)} &
  \makecell{$2727/2589$ \\ ($1.05$)} &
  \makecell{$2786/2589$ \\ ($1.08$)} &
  \makecell{$2787/2589$ \\ ($1.08$)} &
  \makecell{$2761/2589$ \\ ($1.07$)} &
  \makecell{$2719/2588$ \\ ($1.05$)} &
  \makecell{$2722/2588$ \\ ($1.05$)} &
  $\chi^2 / \nu$ &
  \makecell{$2823/2591$\\ ($1.09$)} &
  \makecell{$2755/2591$\\ ($1.06$)} &
  $\chi^2 / \nu$ &
  \makecell{$2736/2586$\\ ($1.06$)} &
  $\chi^2 / \nu$ &
  \makecell{$2720/2588$\\ ($1.05$)} \\
\enddata

\tablecomments{* represents a parameter fixed in the fit. Subscripts d, r and g correspond to the different components of the model: \texttt{diskbb}, \texttt{relxill} (or same family), and \texttt{gaussian}.  For model 7, the numbers reported in parentheses correspond to the \texttt{xillver} component. In the $E_{\rm cut} / kT_e$ row, $E_{\rm cut}$ corresponds to the \texttt{relxill}, \texttt{relxillD}, or \texttt{relxilllp} models, while it corresponds to $kT_e$ to the \texttt{relxillCp} or \texttt{relxilllpCp} models.}
\end{splitdeluxetable*}

\begin{figure}[ht]
\includegraphics[width=0.49\textwidth]{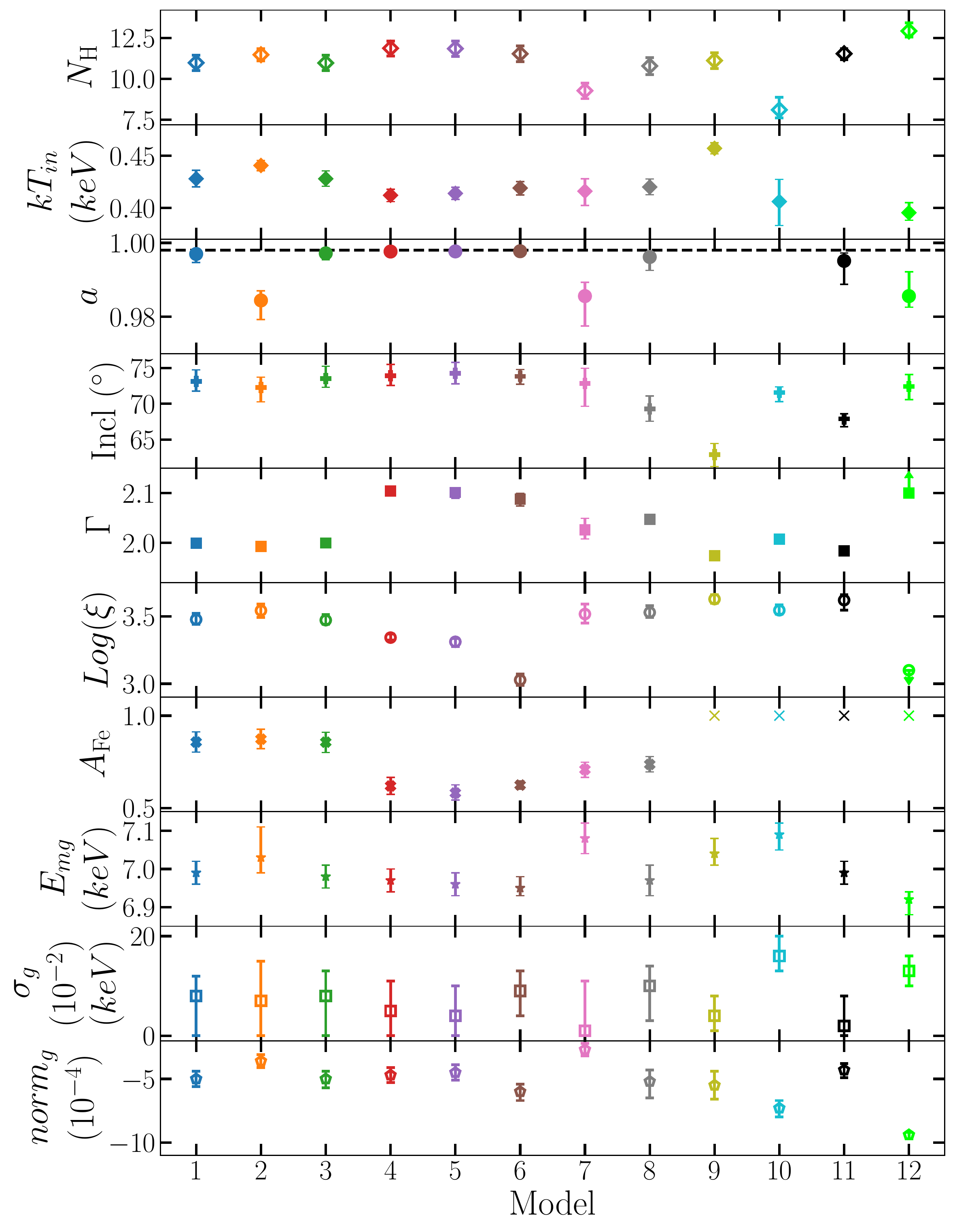}
\caption{Change in the predicted values of seven parameters of interest, across models.  The first panel shows the change in the hydrogen column density ($N_{\rm H}$).  The second panel shows the temperature parameter of the \texttt{diskbb} component, where this component was used.  The third panel shows the measured spin of the black hole, omitting the poor constraints of Models 9 and 10.  The black dashed line represents the upper limit on the spin of a black hole, of 0.998.  The fourth panel shows the inclination of the inner accretion disk.  The fifth and sixth panels show the power law index $\Gamma$ and Log( $\xi$), respectively.  The seventh panel shows the predicted iron abundance, $A_{\rm Fe}$, in solar units. In Models 9, 10, 11, and 12, this was fixed to 1.  The last three panels show the predicted parameters of the \texttt{Gaussian} component of the models.
}
\label{fig:params}
\end{figure}

In order to understand the nature of the inner accretion flow in EXO 1846$-$031 as well as possible, and to further test the very high spin value, we explored fits with a total of 11 other models.  Their results are presented fully in Table \ref{models}.  The variation of some parameters of interest between models is shown in Figure \ref{fig:params}.   The models can be loosely grouped into examinations of the corona and emissivity, the assumed disk density, the role of reflection from the outer disk, and the model used to describe thermal emission from the accretion disk itself: 

\noindent$\bullet$ Models 2 and 3 explore the dependence of the spin results and overall fit on the assumed nature of the hard X-ray corona.  Specifically, Model 2 fixes the outer emissivity index at the Euclidean value of $q_2 = 3$, and Model 3 fixes the breaking radius of the broken emissivity function at $R_{\rm br} = 6 r_{\rm g}$, corresponding to the ISCO for a zero-spin black hole.  

\noindent$\bullet$ Models 4, 5, and 6 test the dependence of the fit on the assumed density of the accretion disk (predicted by \citealt{2016MNRAS.462..751G}).  While models 1, 2, and 3 have a built-in fixed density of log$(n)=15 \: (\rm cm^{-3})$, by switching the \texttt{relxill} component with \texttt{relxillD}, the full density parameter space was probed by fixing log$(n)=15,17,\text{and } 19$ in models 4, 5, and 6.  

\noindent$\bullet$ In other sources, there are indications of warps in the inner or outer disk that may contribute to the overall reflection spectrum (e.g., \citealt{2018ApJ...860L..28M}).  Model 7 therefore tests the possibility that reflection from the outer regions of the disk, with lower ionization, might contribute to the overall reflection spectrum in EXO 1846$-$031.  
This test was implemented by linking the inner and outer emissivity indices ($q_{1} = q_{2}$), and fixing $R_{\rm br}$ to an arbitrary value -- turning the emissivity profile of \texttt{relxill} to an unbroken power law.  For consistency, any distant reflection flux was modeled by adding \texttt{xillver} to the model.  The parameters of \texttt{xillver} in model 7 are linked to those of \texttt{relxill}, with the exception of the ionization parameter and the normalization of the component. 

\noindent$\bullet$ The basic \texttt{relxill} model describes the emisivity of the corona as a power law of index $\Gamma$ with a high-energy cutoff $E_{\rm cut}$.  A more accurate description of the emissivity of a hot corona can be achieved through a thermal Comptonization continuum.  Model 8 replaces the \texttt{relxill} component with \texttt{relxillCp}, which computes the coronal illumination continuum using \texttt{nthComp} (\citealt{1996MNRAS.283..193Z}; \citealt{1999MNRAS.309..561Z}).  The high-energy cutoff $E_{\rm cut}$ parameter is replaced by the electron temperature in the corona $kT_e$. 

\noindent$\bullet$  Whereas allowing the emissivity parameters to vary (e.g., in Models 1-6) essentially allows one to reverse engineer the characteristics of the primary hard X-ray corona, it is also practical to consider the effects of an assumed, self-consistent primary corona.  Models 9 and 10 assume a ``lamppost'' geometry for the corona by replacing \texttt{relxill} with \texttt{relxilllp} and \texttt{relxilllpCp}, which assume the coronal illumination to be described by a cutoff power law and thermal Comptonization, respectively.  The complete models are \texttt{mtable\{nuMLIv1.mod\}*constant* TBabs*(diskbb + relxilllp + gaussian)} and \texttt{mtable\{nuMLIv1.mod\}*constant* TBabs*(diskbb + relxilllpCp + gaussian)}.  Due to the inability of these models to constrain the Fe abundance, this has been fixed to the solar one.

\noindent$\bullet$ The \texttt{diskbb} component is a commonly used model, but one that lacks important physics.  New disk models include key physics, and also permit black hole spin constraints.  In some cases, it has been possible to fit spectra with disk continuum and disk reflection models with linked spin parameters (e.g., \citealt{2009ApJ...697..900M}).  Model 11 therefore also uses \texttt{relxill}, but replaces a highly simplified treatment of the accretion disk with a relativistic one, by replacing \texttt{diskbb} with \texttt{kerrbb} (\citealt{2005ApJS..157..335L}).  The spin and inclination parameters in \texttt{kerrbb} were tied to the ones in \texttt{relxill}.  The distance to EXO 1846$-$031 was fixed at the value of 7~kpc estimated by \citealt{1993A&A...279..179P}.  This checks whether the fit achieved with Model 1 is affected by the unphysical assumptions of the \texttt{diskbb} component.  However, Model 11 offers an additional benefit: if the black hole distance and spin are known, it is nominally possible to constrain the mass of the black hole using \texttt{kerrbb}. Similar to Models 9 and 10, the Fe abundance is fixed to unity in this model.

\noindent$\bullet$ While Models 4--6 probe densities between log$ (n)=15$ and log$ (n)=19$, even higher densities can be tested by changing the \texttt{relxill} components to \texttt{reflionx} (\citealt{2005MNRAS.358..211R}).  In particular, we used the high-density model \texttt{reflionx$\_$hd} produced by \citealt{2018ApJ...855....3T} using the code provided by \citealt{2007MNRAS.381.1697R}.  We replaced the \texttt{relxill} component of our baseline model with \texttt{relconv(cutoffpl+reflionx$\_$hd)}, where the \texttt{reflionx$\_$hd} describes the reflection component of the spectrum and \texttt{cutoffpl} describes the direct component.  These two are convolved with  relativistic accretion disk line profiles using \texttt{relconv} (\citealt{2010MNRAS.409.1534D}).  The \texttt{reflionx$\_hd$} model was built to have fixed $A_{\rm Fe}=1$ and $E_{\rm cut}=300~keV$.  The complete Model 12 is \texttt{mtable$\{$nuMLIv1.mod$\}$*constant*TBabs*(diskbb + relconv(cutoffpl + reflionx$\_$hd) + gaussian)}.  

The parameters of the fits are highly consistent between all of the models.  The power law index is close to $\Gamma = 2$ except for Models 4--6, where it is only $\Delta\Gamma = 0.1$ steeper, and for Model 12, where $\Delta\Gamma = 0.3$.  Where the variant of \texttt{relxill} allowed the high-energy cutoff to vary, it remained high in all cases.  Models 4--6 fix this at 300~keV, producing slightly worse fits when compared to our baseline model.  This again is suggesting a hybrid and likely compact corona, as a high temperature requires a small size for gas to be bound to the black hole.    In all cases, the disk temperature remains low, $kT \leq 0.46$~keV.   The Fe abundance is sub-solar in all cases, ranging between 0.58 and 0.87.  The estimated ionization of the disk (log $\xi$) takes large values, between $\sim3.30-3.60$, broadly consistent with recent prior results using similar models.  The lowest ionization values (log $\xi = 3.03$ and log $\xi = 2.21$) are found when the highest disk density is assumed (log $n = 19$ in Model 6 and log $n = 20$ in Model 12).  Additionally, the Gaussian component of the models remains significant with at least a $4\sigma$ level of confidence.

Model 8 changes the emissivity of the corona, describing it through the more physical model of a thermal Comptonization continuum as opposed to a power law with a high-energy cutoff.  The fit is equally good as our baseline model and all the common parameters are highly consistent between the two models.  The high measured value of the electron temperature once again suggests the presence of a hot, compact corona.  \citealt{2001MNRAS.321..549M} use energetics arguments to show that the size of black hole coronae powered by hot electrons has to be on the order of $10^3~r_g$ in order to explain the observed hard X-ray fluxes. For the most ideal choice of parameters, given the electron temperature inferred by Model 8, the minimum size of the corona has to be at least $\sim 200 ~r_g$. This large physical size of the corona conflicts with the observed variability timescales in black holes (see e.g., \citealt{1999MNRAS.306L..31P}; \citealt{2000MNRAS.318..857L}; \citealt{2015MNRAS.451.4375F}).   \citealt{2001MNRAS.321..549M} argue that in order to account for the sizes inferred through observations, the energy in the corona must be stored as magnetic fields generated by a sheared rotator.  These arguments suggest that despite this model producing a good fit, the thermal Comptonization continuum models might not offer a good physical explanation of the processes driving this system. 

However, this does not exclude the possibility that the coronal emission is caused by the very compact base of a jet and a more extended electron-dominated thermal corona.  This configuration of a corona with a strong central concentration and a more extended component can explain the low values predicted for the outer emissivity index $q_2$.  \citealt{2012MNRAS.424.1284W} predict that the emissivity expected close to a rapidly spinning black hole is better approximated by a steep emisivity index $q_1$ at small radii, followed by an almost flat emissivity up to the outer regions of the disk, where the emissivity becomes constant, taking the Euclidean value of $q_3=3$. Current models do not yet have this functionality. 

The most important outcome of these additional models is that the spin determination remains very high in all cases apart from Models 9 and 10, which measure $a = 0.2^{+0.5}_{-0.6}$ and $a = 0.1^{+0.7}_{-0.7}$ suggesting that the ``lamppost'' model is unable to constrain the spin of this object.  The height of the corona ($\sim24 \: r_{\rm g}$ and $\sim69 \: r_{\rm g}$) predicted by Models 9 and 10 makes it unable to accurately measure spin, as explained by \citealt{2014MNRAS.439.2307F}, \citealt{2017ApJ...851...57C}, or \citealt{2018A&A...614A..44K}.  Similarly, the inclination remains close to $\theta = 70^{\circ}$ in all cases apart from Model 9 ($\theta = 63^{\circ}\pm 2^{\circ}$). The top right panel of Figure \ref{fig:hist} shows the spin and inclination measurements for the different models that were tested.  Model 9 yields a significantly worse fit than our baseline model; indeed, it is the worst fit of any of the models considered, with $\Delta\chi^2=97$ worse for three extra degrees of freedom, so it was not included in the figure.  Our baseline model (Model 1) is the overall best fit, in terms of the reduced $\chi^{2}$ statistic.  The addition of distant reflection from the outer accretion disk is therefore not required by the data, and there is no evidence for or against a flared or warped disk in EXO 1846$-$031.

Model 11 strengthens the predictions of our baseline model in that it measures consistent values of all of the reflection parameters; in this sense, the spin determination is not only independent of the coronal properties accounted for by our models, but also independent of the disk model that is assumed.  For the distance of 7~kpc estimated by \citealt{1993A&A...279..179P}, \texttt{kererbb} provides an estimate of the mass of the central black hole of $9\pm 5 M_\odot$ ($1\sigma$ errors).  This value cannot be considered as definitive due to the limited low energy sensitivity of \textit{NuSTAR} and due to the degeneracy between the parameters of the \texttt{kerrbb} model.  Further measurements of the blackbody continuum spectrum are required for better constraints on the mass of the black hole, taking advantage of the determination of the relativistic reflection component of the spectrum of this work. 

Replacing the \texttt{relxill} component of our baseline model with the high density version of \texttt{reflionx} produces an equally good fit.  However, while the preferred disk density of log$ (n)=20$ is higher than the ones probed by \texttt{relxillD}, this parameter appears to be highly degenerate with the disk ionization $\xi$ and the normalization of the component.  Nevertheless, the \texttt{diskbb} temperature, emissivity profile, spin of black hole and inclination of the inner accretion disk are well constrained and consistent with the predictions of the other models.

\section{Discussion} \label{sec:disc}

We observed the black hole candidate EXO 1846$-$031 with {\it NuSTAR} during its outburst in late 2019.  Extremely sensitive spectra were obtained.  Fits to these spectra suggest that the source was likely observed in a hard intermediate state.  The most interesting and important feature of the spectra is a strong, highly skewed relativistic disk reflection spectrum.  Our preferred or ``baseline'' spectral model measures a very high black hole spin parameter, $a=0.997_{-0.002}^{+0.001}$, via the well-known \texttt{relxill} model, with the value being larger than $a=0.984$ at a $5\sigma$ confidence.  
We explored a series of 12 models in total, allowing for different treatments of the corona, different disk continua, different values of the accretion disk density, and the possibility of additional reflection from the outer accretion disk.  In every case, a very high spin parameter results, indicating that our key result is particularly robust.  Similarly, careful examinations of potential parameter degeneracies in all models find that the spin and inclination are both well determined, again indicating a robust spin measurement.  Here, we examine some additional strengths and limitations of our analysis, and attempt to place our results into a broader context.

The black hole spin parameter implied by our spectral models is extremely high.  Indeed, the values obtained are at the absolute limit that is theoretically possible.  A black hole cannot have a spin above $a = 1$ without violating causality, but other pragmatic considerations point to a lower functional limit.  A rapidly spinning black hole will tend to capture photons with negative angular momentum, leading to a limit of $a \leq 0.998$ (\citealt{1974ApJ...191..507T}).  
Our results nominally push to this limit.  However, we are only able to report {\it statistical} errors in this work.  We have endeavored to understand the systematic errors incurred through our choice of models; these appear to be comparable to the statistical errors.  However, the dominant source of systematic error is likely tied to how closely the accretion disk adheres to the test particle ISCO.  If the gas density does not fall rapidly inside of the ISCO, then the material may still emit thermal radiation and may still be reflective, biasing spin measurements toward high values.  In this case, systematic errors might allow for a lower spin parameter, comfortably below the limit set by \citealt{1974ApJ...191..507T}.

Numerical simulations suggest that the disk obeys the test particle ISCO for $L/L_{\rm Edd} \leq 0.3$ (see, e.g., \citealt{2018ApJ...857....1F}).  The distance to EXO 1846$-$031 was estimated to be 7~kpc by assuming that the flux observed during its 1985 outburst corresponds to a luminosity of $10^{38}~erg/s$ (\citealt{1993A&A...279..179P}), and we have estimated the mass of the black hole to be $M = 9\pm 5~M_{\odot}$.  Our baseline model gives an absorbed flux of $F = 1.0 \: (1.1) \times 10^{-8}~{\rm erg}~{\rm cm}^{-2}~{\rm s}^{-1}$ in the 3--79 (0.5--100)~keV band.  This corresponds to an unabsorbed flux $F = 1.2 \: (4.0)\times 10^{-8}~{\rm erg}~{\rm cm}^{-2}~{\rm s}^{-1}$ in the 3--79 (0.5--100)~keV band.  These values imply Eddington fractions of 0.06 and 0.25, respectively, and are within the limit where disks are expected to obey the test particle ISCO.  We note that for the assumed distance, the black hole mass could be as low as the theoretical upper limit for neutron star masses, and still be within the range wherein disks are expected to obey the test particle ISCO.  

Spin parameters as high as that measured in EXO~1846$-$031 may nominally serve as a discriminant between black holes and neutron stars in X-ray binaries wherein the primary mass is unknown.  Type-I X-ray bursts, kHz QPOs, and other characteristic phenomena have not been detected in EXO 1846$-$031, for instance.  This is merely an absence of evidence for a neutron star; this is not evidence {\it against} a neutron star nor evidence {\it for} a black hole.  However, very simple considerations demand that a neutron star will break up for $a \geq 0.7$ (\citealt{2011ApJ...728...12L}).   In this sense, the spin measurement that we have obtained in EXO 1846$-$031 {\it requires} that the primary is a black hole.  

A black hole with an initial spin of $a = 0$ must accrete at least half its mass to reach a spin parameter of $a = 0.84$ (\citealt{1970Natur.226...64B}).  Since stellar-mass black holes in LMXBs are more massive than their companion stars, and since very massive stars have extremely short life spans, it is unlikely that stellar-mass black holes can significantly alter their spin parameters.  The very high spin parameter we measure in EXO 1846$-$031 suggests that the black hole was born with a very high spin.  Simulations of stellar collapse have found that while most black holes are born with slow rotation rates (e.g., \citealt{Fuller_2019}), spin parameters as high as $a = 0.9$ are possible (see e.g., \citealt{Heger_2005}).  A black hole would need to accrete $\sim33\%$ of its initial mass to increase its spin from $a = 0.9$ to $a = 0.98$ (based on the prescription of \citealt{2008ApJ...682..474B}).  This implies that the spin of the black hole in EXO 1846$-$031 may require a fortuitous combination of parameters and effects, potentially including an especially massive progenitor star, a low progenitor metallicity that prevents extensive mass loss, solid-body rotation within the star that effectively stores angular momentum, and a collapse that does not translate the angular momentum into a kick.

The mean of the distribution of black hole spins measured through the relativistic reflection method has been estimated to be around $a=0.66$  (\citealt{2015PhR...548....1M}). Still, very high spin values have previously been measured, such as $a=0.998_{-0.009}$ for GS 1354-645 (\citealt{2016ApJ...826L..12E}), or $a=0.985_{-0.014}^{+0.005}$ for 4U 1630-472 (\citealt{King_2014}). In the case of GRS 1915+105, relativistic disk reflection suggests a spin of $a=0.98^{+0.01}_{-0.01}$ (\citealt{2009ApJ...706...60B}; \citealt{2013ApJ...775L..45M}), results preceded by the measurement of \citealt{2006ApJ...652..518M} that predicted $a>0.98$ by using the disk continuum fitting method. 

Gravitational-wave observations of inspiral, coalescence, and ringdown of BBH systems allow for mass and spin estimates of the components both before and after the merger. \citealt{2019ApJ...882L..24A} estimates that half of the black holes in black hole binary systems have spins less than 0.27, and 90\% have spins less than 0.55.  The first three gravitational-wave events detected have been inferred to have a premerger spin of the primary black hole $a\leq0.7$ for GW150914 and GW151012 and $a\leq0.8$ for GW151226, while the final black hole spins are around 0.7 in all three cases (\citealt{PhysRevX.6.041015}).  Additionally, \citealt{2020arXiv200407382B} finds that based on the first two LIGO observing runs, the observed black hole population is consistent with a low natal spin population for premerger black holes and inconsistent with a uniform or high spin population. Similar results were obtained by \citealt{2018MNRAS.473.4174Z} and \citealt{2020A&A...635A..97B}.  Still, premerger black hole spins are strongly dependent on the analysis pipeline and choices of priors for the Bayesian analysis (\citealt{2017PhRvL.119y1103V}), with some techniques predicting high spin values (e.g., \citealt{2019arXiv191009528Z}, \citealt{2019PhRvD.100b3007Z}). 

One particularly interesting gravitational wave-detected BBH merger is GW190412, for which \citealt{2020arXiv200408342T} predict the coalescence of a $\sim30M_\odot$ black hole with spin $a=0.43^{+0.16}_{-0.26}$ (90\% confidence) and an $\sim8M_\odot$ secondary.  It was argued by \citealt{2020ApJ...895L..28M} that an alternative explanation for the signal is a merger of a near spinless primary black hole with a low-mass secondary with a high spin value ($a=0.88^{+0.11}_{-0.24}$), aligned with the orbital angular momentum.  Later, \citealt{2020ApJ...899L..17Z} argued that the data prefers the intermediate spin value of the primary black hole.  While current gravitational-wave measurements make it difficult to distinguish between these models, understanding the distribution of spins in the black hole population in accreting stellar-mass black holes can provide insight into the formation mechanism of these populations and yield improved priors for both current LIGO and future GW observatory detections.

Finally, the {\em NuSTAR} spectra require the addition of a Gaussian line to model a feature that is likely a H-like Fe XXVI absorption line.  The detection of Fe XXVI without significant He-like Fe XXV requires that the gas be very highly ionized.  Disk winds with high ionizations are preferentially detected in systems viewed at high inclination (e.g., \citealt{2012MNRAS.422L..11P}); given the high inclinations we have measured in EXO 1846$-$031, it is likely that this feature arises in a disk wind.  However, the feature is not measured to be at exactly the same energy with every model, and some measurements carry errors that also make it unclear if the absorbing gas is truly outflowing.  In the near future, pairing {\it NuSTAR} with {\it XRISM} (\citealt{2018SPIE10699E..22T}) will make it possible to study relativistic reflection and winds simultaneously.

\acknowledgements
We thank the {\em NuSTAR} director, Fiona Harrison, and the mission scheduling team for making this observation, and Michael Parker for constructing the \texttt{reflionx$\_$hd} table model.
This research has made use of data and software provided by the high-energy Astrophysics Science Archive Research Center (HEASARC), which is a service of the Astrophysics Science Division at NASA/GSFC, and of the NuSTAR Data Analysis Software (NuSTARDAS) jointly developed by the ASI Science Data Center (ASDC, Italy) and the California Institute of Technology (Caltech, USA).
John A. Tomsick acknowledges partial support from NASA ADAP grant 80NSSC19K0586.
We also thank the anonymous referee for their comments and suggestions, which have clarified and improved this paper.

\newpage

\bibliography{paper}{}
\bibliographystyle{aasjournal}

\appendix

\section{Corner plot}\label{sec:app}

\begin{figure}[ht]
\includegraphics[width=0.95\textwidth]{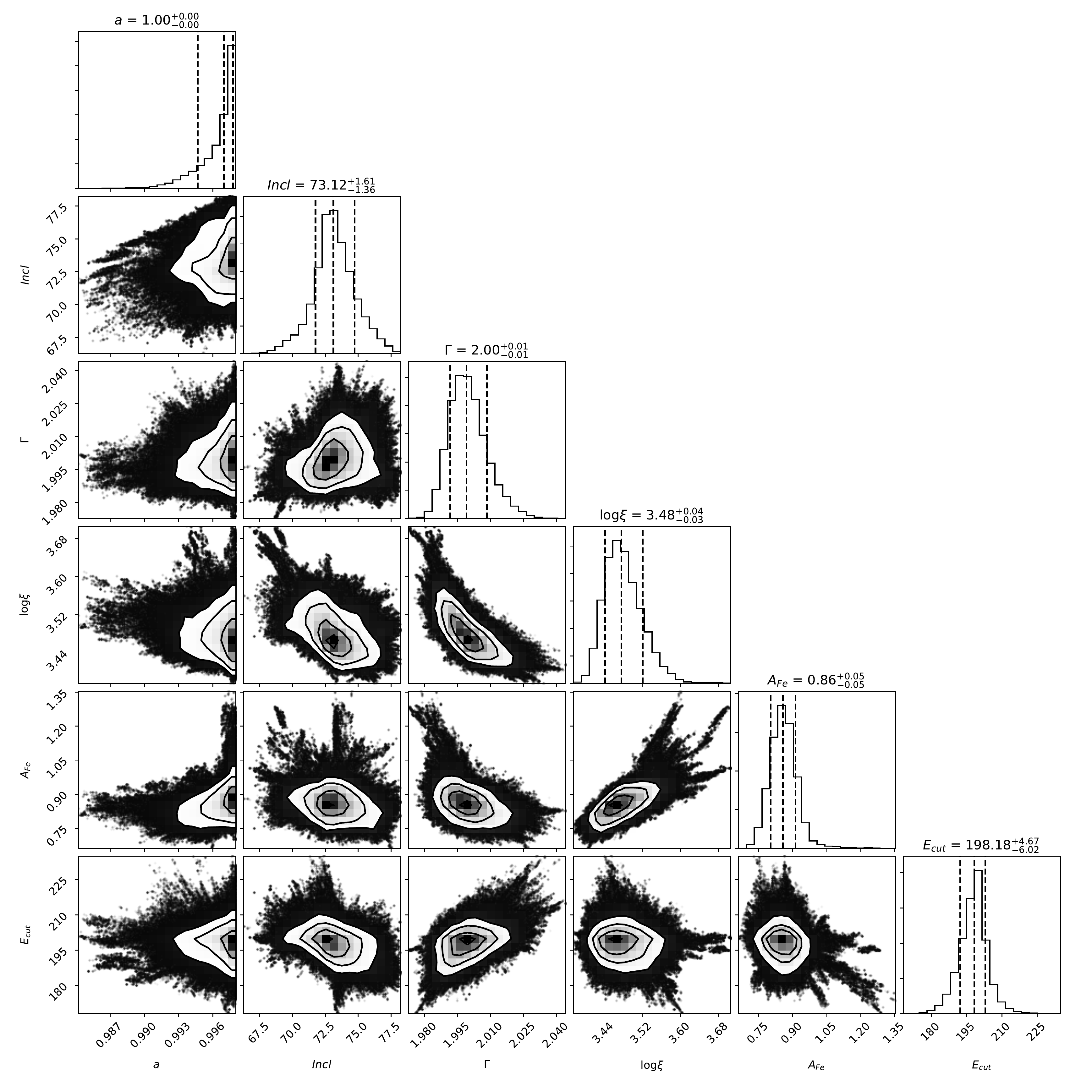}
\caption{Corner plot of the reflection component parameters in the baseline \texttt{mtable\{nuMLIv1.mod\}*constant*TBabs* (diskbb+relxill+gaussian)} model.}
\label{fig:corner}
\end{figure}

\end{document}